\documentclass[letterpaper]{article} 
\usepackage[]{aaai25}  
\usepackage{times}  
\usepackage{helvet}  
\usepackage{courier}  
\usepackage[hyphens]{url}  
\usepackage{graphicx} 
\usepackage{natbib}  
\usepackage{caption} 
\usepackage{algorithm}
\usepackage{algorithmic}
\usepackage{amsmath}
\usepackage{booktabs}
\usepackage{xcolor}

\usepackage{algorithm}
\usepackage{algorithmic}
\usepackage{amsmath}
\usepackage{booktabs}

\usepackage{enumitem}
\usepackage{xcolor}
\usepackage{booktabs}
\usepackage{tabularx}
\usepackage{tcolorbox}
\usepackage{cuted}

\usepackage{newfloat}
\usepackage{listings}
\DeclareCaptionStyle{ruled}{labelfont=normalfont,labelsep=colon,strut=off} 
\floatstyle{ruled}
\newfloat{listing}{tb}{lst}{}
\floatname{listing}{Listing}
\title{Explaining the Reputational Risks of AI-Mediated Communication: Messages labeled as AI-assisted are viewed as less diagnostic of the sender’s moral character}
\author{
    Pranav Khadpe\equalcontrib,
    Kimi Wenzel\equalcontrib,
    George Loewenstein,
    Geoff Kaufman
}
\affiliations{

    Carnegie Mellon University\\
    \{pkhadpe, kwenzel, gl20, gfk\}@andrew.cmu.edu
}

\usepackage{bibentry}

\begin{document}

\maketitle

\begin{abstract}
When someone sends us a thoughtful message, we naturally form judgments about their character. But what happens when that message carries a label indicating it was written with the help of AI? This paper investigates how the appearance of AI assistance affects our perceptions of message senders. Adding nuance to previous research, through two studies ($N=399$) featuring vignette scenarios, we find that AI-assistance labels don't necessarily make people view senders negatively. Rather, they dampen the strength of character signals in communication. We show that when someone sends a warmth-signalling message (like thanking or apologizing) without AI help, people more strongly categorize the sender as warm. At the same time, when someone sends a coldness-signalling message (like bragging or blaming) without assistance, people more confidently categorize them as cold. Interestingly, AI labels weaken both these associations: An AI-assisted apology makes the sender appear less warm than if they had written it themselves, and an AI-assisted blame makes the sender appear less cold than if they had composed it independently. This supports our \textit{signal diagnosticity explanation}: messages labeled as AI-assisted are viewed as \textit{less diagnostic} than messages which seem unassisted. We discuss how our findings shed light on the causal origins of previously reported observations in AI-Mediated Communication.
\end{abstract}

%

\section{Introduction}
How people communicate is one of the fundamental signals we use to judge their moral character. These judgments affect whom we choose to approach and whom we choose to trust~\cite{fiske2018model}. Many of us instinctively trust the colleague who remembers to thank the cleaning staff unprompted each day, while questioning the warmth of the manager who thanks his staff only when reminded by HR. As society adapts to communication that is assisted by artificial intelligence (AI)---whether speeding up typing with autocomplete, streamlining exchanges with autoreply, or delegating thoughtfulness to ChatGPT---questions about how it will affect interpersonal relationships have become a matter of public debate and a focus of scholarly research~\cite{jakesch2019ai,liu2022will, weiss2022effects, hohenstein2020ai}: if you received a thank you note labeled ``written with the help of AI,'' would you perceive the sender as warmer or colder?  

Empirical work on these perceptions consistently finds that when warmth-signalling messages such as greetings, thanks, and apologies are labeled as AI-assisted, they are associated with lower ratings of the sender's moral character (reduced perceived warmth) compared to identical messages presented without AI labels~\footnote{\textit{Warmth} is the standard measure of moral character in social cognition literature~\cite{fiske2007universal}.}~\cite[e.g.][]{jakesch2019ai,liu2022will, weiss2022effects, hohenstein2020ai, hohenstein2023artificial, mieczkowski2021ai}. However, few explanatory mechanisms have been proposed for why this is the case. Two distinct mechanisms could explain this disparity in warmth judgments: either people view the AI-assisted message as negative evidence of warmth (as evidence that the sender is actually cold and calculated), or they see it as merely weaker positive evidence of warmth (accepting that the sender possesses warmth, but not intensely enough that its verbal expression flows without assistance).

In this paper, we describe, test, and ultimately find support for the \textit{signal diagnosticity explanation}, which is that appearance of AI assistance makes messages weaker (less diagnostic) evidence rather than negative evidence. The implication of this is that AI labels \textit{dampen} the effect a message would have otherwise had (as opposed to AI labels having a categorical negative effect). The sender of a thanks or apology is seen as less warm when AI-assisted, and the sender of brags and blames is seen as less cold (more warm) when AI-assisted. Our two studies provide evidence for this explanation by examining how people perceive senders when AI labels are attached to both warmth-signalling messages (thanks and apologies) and coldness-signalling messages (brags and blames). Study 1 directly tested message diagnosticity: apology messages without AI labels were viewed as more diagnostic of warmth than those with AI labels, while blame messages without AI labels were viewed as more diagnostic of coldness than those with AI labels. Study 2 tested effects on actual warmth judgments. AI-assisted thanks and apologies received significantly lower warmth ratings than non-AI-assisted messages. Yet strikingly, this drop disappeared for brags and blames, supporting the idea that AI labels have a dampening effect rather than a categorical negative effect. 

By placing phenomena observed in AI-Mediated Communication (AIMC) on firm psychological foundations, we discuss how our work sheds light on previously reported phenomena, including how AIMC affects interpersonal trust~\cite{jakesch2019ai} and serves as a moral crumple zone~\cite{hohenstein2020ai}.  

\subsubsection{Contributions:} Our contributions are (1) an explanation for how appearance of AI assistance affects judgments of a message sender's moral character; (2) empirical support for the explanation across two studies; (3) tracing previously reported empirical phenomena in AIMC to their psychological origins.

\section{Approach}
\subsubsection{The Signal Diagnosticity Explanation:} The diagnosticity approach~\cite{skowronski1987social} frames character judgments as a probabilistic categorization process. When evaluating others, people analyze signals---like the messages someone sends---to determine whether that person belongs in the ``warm'' or ``cold'' category. Signals contribute to categorization probabilistically rather than with absolute certainty~\cite{skowronski1987social}. For instance, someone who sends a thoughtful message of thanks probably belongs in the warm category, although there is a real, though lesser, probability that they belong to the cold category. 

Signals are defined as more diagnostic when they lead to higher perceived probabilities that a person belongs to one category (e.g. warm) and lower perceived probabilities that the person belongs to the alternate category (e.g. cold). Therefore, more diagnostic signals lead to more confident assignment of a person to either the warm or cold category.

Our signal diagnosticity explanation proposes that messages without AI labels are more diagnostic because people intuitively believe that genuinely warm individuals express warmth directly (through thanks or apologies) with greater ease, while genuinely cold individuals express coldness directly (through brags or blame) more naturally. We propose that when thanks and apologies appear unassisted, they serve as stronger probabilistic evidence that the sender belongs in the warm category. Similarly, when brags and blames appear unassisted, they function as stronger probabilistic evidence that the sender belongs in the cold category. We suggest (and show) that AI labels reduce this diagnosticity in both cases: AI-assisted thanks and apologies become weaker evidence that the sender belongs in the warm category, and AI-assisted brags and blames become weaker evidence that the sender belongs in the cold category.

\subsubsection{The Current Work:}
In both studies, participants read a fictional scenario involving interactions between them and a fictional character. The scenario was manipulated across conditions to create warrant for the character to thank, blame, apologize to, or brag to the participant. Then, participants read a message from the character either thanking, apologizing, bragging or blaming, presented with or without an AI label (``parts of this message were written with the help of AI''). Participants then provided their judgments of the message and the fictional character. In Study 1, we focused only on an apologies and blames, and we elicited subjective probabilities from participants to compute message diagnosticity. In Study 2, we expanded the set of messages to also include thanks and brags, and we captured participants judgments of the sender's warmth, the sender's competence, and open-ended responses about their attributions of the sender's use of AI. This served as a confirmatory test of the explanation and provided richer insights into the basis of people's judgments. 

\section{Related Work}
We begin by establishing the theoretical foundation for our investigation, drawing on research in social cognition that identifies warmth as a primary dimension of interpersonal judgment. We then make a distinction between judgments of messages versus their senders, before reviewing empirical work on AIMC that motivates our explanatory account.

\subsection{Warmth: A Judgment of Moral Character}
Research in social cognition shows that to navigate our social world safely and effectively, we evaluate others along two fundamental dimensions: warmth and competence~\cite{fiske2007universal, fiske2018model}. \textit{Warmth} captures whether someone intends to help or harm us---distinguishing between those we perceive as warm (trustworthy, kind) versus cold (untrustworthy, hostile). It is typically understood as a judgment of moral character, encompassing traits such as trustworthiness, kindness, and friendliness,~\cite{fiske2007universal, russell2008s}. Meanwhile, \textit{competence} captures whether they have the ability to act on those intentions---their perceived skill and effectiveness~\cite{fiske2007universal, russell2008s}. These two dimensions are shown to emerge reliably across stimuli, culture, and time~\cite{fiske2007universal, russell2008s}.

Of the two dimensions, judgments along the warmth dimension are primary. Warmth is judged first~\cite{willis2006first} and has a larger impact on individual's affective and behavioral reactions \cite{hack2013warmth, fiske2007universal, abele2014communal}. A common explanation for this prioritization is that, from an evolutionary perspective, another person's disposition to help or harm has more consequential implications for one's survival than their ability to act on those intentions \cite{abele2007agency, fiske2007universal, cuddy2008warmth}.

Our investigations focus on the warmth dimension. This is due to two intersecting reasons. First, the fact that warmth judgments are primary and consequential makes it an important dimension of study; most prior work on AIMC studies warmth-related traits~\cite{jakesch2019ai, hohenstein2020ai, hohenstein2023artificial, liu2022will}. Second, focusing on the broader dimension of warmth, rather than individual traits like friendliness or trustworthiness, allows us to explain findings across studies that examine different warmth-related characteristics. This unification provides a foundation for theory-building because while many prior studies on AIMC measure individual warmth-related traits~\cite[e.g., trustworthiness in][]{jakesch2019ai}, few explicitly reference the broader warmth construct~\cite{weiss2022effects, mieczkowski2021ai}, which has limited theoretical synthesis across individual studies.

\subsection{Dissociated Responses to Message and Source}
Character judgments---warmth judgments---are not assessments of the \textit{message} per se; rather they are inferences about the \textit{sender} drawn from the message. This distinction is important because character inferences about the source of communication can often dissociate from reactions to the message itself~\cite{malle2021moral}. For example, you might find a friend's hastily written, typo-filled text message annoying or unclear while still viewing them as a caring and trustworthy person. Dissociated responses are, in fact, known to occur when social action is mediated by AI systems~\cite{renier2021err, bower2021perceptions, liu2022bayesian}. For instance, people are just as persuaded by argument messages presented as AI-generated as they are when presented as human-written, even though they rate the AI source as less trustworthy than a human source~\cite{aydin2024dissociated, gallegos2025labeling}. 

Our focus in this paper is not on how people judge AI-assisted messages, but rather how they judge the \textit{sender} of such messages. While there is research documenting that people view AI-assisted messages as morally inappropriate~\cite {moradi2024fountain, liu2022will}, the primary focus of our explanatory account and empirical investigation is on the inferences people make about the sender upon receiving the message. Focusing on how recipients judge senders helps develop a richer understanding of how AI reshapes interpersonal relationships and trust dynamics.

\subsection{Empirical Investigations of AIMC}
The term AI-mediated communication (AIMC) covers different kinds of mediated communication between people ``in which a computational agent operates on behalf of a communicator by modifying, augmenting, or generating messages to accomplish communication or interpersonal goals'' \cite{hancock2020ai, jakesch2019ai}. Instances of AIMC can be found directly embedded in communication tools (e.g. smart/auto reply\footnote{https://blog.google/products/gmail/save-time-with-smart-reply-in-gmail/} and writing support\footnote{https://www.zdnet.com/article/you-can-now-use-gemini-in-google-messages-if-youre-among-the-lucky-few/}) and can also be facilitated through third-party tools (e.g. drafting a message using a consumer-facing LLM, such as ChatGPT).

Most empirical investigations into the impact of AI mediation on interpersonal communication can be classified into two broad threads of work: a first thread studying \textit{recipient judgments}, and a second thread studying \textit{sender behaviors}. The first thread---and the thread we contribute to---investigates how awareness that a sender used AI affects recipients' judgments~\cite{jakesch2019ai, hohenstein2020ai, hohenstein2023artificial, liu2022will, purcell2023fears, cheong2025role}. The second thread focuses on how AI assistance alters the sender's messaging behavior, including language use~\cite{arnold2020predictive, jakesch2023co, hohenstein2023artificial, mieczkowski2021ai, agarwal2025ai} and communication frequency~\cite{hohenstein2023artificial}.

In this paper, we focus on the first thread: we study the recipient's judgments of the sender, when the message is known to be written with AI. Studies examining these perceptions reveal a consistent pattern: senders of warmth-signalling messages like greetings, thanks, and apologies receive lower warmth ratings when their messages carry AI assistance labels compared to when identical messages appear unassisted~\cite{jakesch2019ai,liu2022will, weiss2022effects, hohenstein2020ai, hohenstein2023artificial, mieczkowski2021ai}. For example, \citeauthor{jakesch2019ai} found that Airbnb hosts are perceived as less trustworthy when participants knew the host used AI~\cite{jakesch2019ai} to write their profile greetings. Similarly, \citeauthor{liu2022will} found that senders of invitations, and condolences---messages that by themselves would enhance perceptions of the sender's warmth---lead to lower warmth judgments when they appear AI-assisted~\cite{liu2022will}. This pattern has been observed across both individual messages~\cite{jakesch2019ai, weiss2022effects, liu2022will} and extended communication episodes~\cite{mieczkowski2021ai, hohenstein2020ai, hohenstein2023artificial}.

Despite this consistent finding, few mechanisms have been proposed to explain why perceptions of AI assistance reduce warmth judgments in the above cases. Thus, we extend prior work by proposing that AI-labeled messages are viewed as \textit{less diagnostic} of the sender's moral character. Crucially, our explanation also predicts when AI labels will \textbf{not} lower warmth judgments: for coldness-signalling messages like brags and blames. We provide empirical support for this explanation across the studies we present next.

\section{Study 1}
Our studies were designed to investigate whether the presence of AI labels affects message diagnosticity (Study 1), and to investigate the effect of AI labels on downstream judgments of the sender's warmth (Study 2). In the first study, participants rated the subjective probabilities that a particular message is associated with the trait category warm, and with the opposite trait category: cold. From this data, we computed perceptions of message diagnosticity for warmth and coldness. This study featured a 2 (message type: apologizing vs. blaming) $\times$ 2 (authorship label: AI label vs. no label) between-subjects design. Our studies used the method of hypothetical scenarios. The method is commonly used in studies of interpersonal perception~\cite{caprariello2009social, chaudhry2019thanking}, including those focused on AIMC~\cite{liu2022will, jakesch2019ai}. Hypothetical scenarios allow us to precisely control the variables of the setup---message type and authorship label---that are the focus of our investigation.   

\subsection{Method}

\subsubsection{Participants:} Using the online crowdsourcing platform \textit{Prolific}, we recruited 130 participants. This sample goal was based on a power analysis to detect a medium effect of size $f \geq 0.25$ (powered at 80\% with a significance level of 0.05). After excluding participants who failed the attention check, or the comprehension check, we were left with 119 participants for analysis (29-30 in each condition; mean age = 40.8; 60 women and 59 men). Participants received \$2 for completing the 10 minute survey.    

\subsubsection{Stimulus Development:} We wanted to test how AI labels affect diagnosticity for both warmth-signalling (e.g. thanks and apologies) messages and coldness-signalling messages (e.g. brags and blames). Drawing on prior work that has established how apologies are warmth-signalling and blames are cold-signalling~\cite{chaudhry2019thanking}, in this study we focus on apologies and blames are representative of those two categories. Specifically, we follow previous work~\cite{chaudhry2019thanking} and operationalize an apology message as the sender ``\textit{taking} blame for a negative outcome'', and blame message as the sender ``\textit{giving} the recipient blame for a negative outcome''.

To create social warrant for the participant to be receiving the message, we used a hypothetical scenario involving the participant and a friend (``John'') who are members of a community organization planning neighborhood events. In the scenario, one character helps the other plan an event, which then receives harsh criticism at a community meeting. We manipulated who was the primary organizer versus helper to create contexts where either an apology or blame message from John would be situationally appropriate: an apology was warranted when the participant organized and John helped, while blame was warranted when John organized and the participant helped.

Both messages were 3-4 sentences long and generated using ChatGPT (GPT-4). The prompt included the scenario corresponding to the message, and asked the system to generate a short, 3-4 sentence long, text message from John. For participants assigned to see the message with AI labels, the message was additionally appended with a label that read ``parts of this message were written with the help of AI''. This resulted in four conditions: apology without label, apology with AI label, blame without label, and blame with AI label. (The scenario, the message content, and how they were shown to participants can be found in Supplementary Material.) 

\subsubsection{Trait-Behavior Probability Scale:}
\label{scale}
To reason about the diagnosticity of each of the four messages, we need estimates of the perceived probabilities that the message is associated with the trait category \textit{warm}, and with the opposite category: \textit{cold}. For this, we draw on the trait-behavior probability scale~\cite{skowronski1987social} used to measure diagnosticity. In the scale, a particular behavior is paired with opposing traits in sentences of the form: "Would a (trait) person ever (behavior)?". For example, in Skowronski and Carlson's own study---which measured the diagnosticity of different behaviors for the trait of honesty (and dishonesty)---a pair of items shown to an individual participant were: ``Would an honest person ever steal money from his roommate's wallet?'' and  ``Would a dishonest person ever steal money from his roommates wallet?''. 

Participants respond to these questions on a 9-point scale, ranging from (1) \textit{extremely unlikely to perform the behavior} to (9) \textit{extremely likely to perform the behavior}, with (5) \textit{moderately likely
to perform the behavior} at the midpoint.

To reason about diagnosticity of a behavior, the responses are used to calculate a signal-validity score. This is calculated as a ratio of the perceived probabilities: ``The numerator is the rated probability
that an actor with a given trait will perform a specified behavior, and the denominator is the sum of this probability with the rated probability that an actor with the opposing trait will perform the same behavior.'' In the above example, if a participant rates the probability that an honest person would steal the money at $1$, and the probability that a dishonest person would steal the money at $6$, then the signal-validity of the behavior \textit{steal money from roommates wallet} for the trait category of \textit{honest} would be $0.14 \text{ } (1/[1+6])$, and the signal-validity of this behavior for the category \textit{dishonest} would be $0.86 \text{ } (6/[1+6])$. We adapt the trait-behavior probability scale as we describe next.

\subsubsection{Procedure and Measures:}
Participants were randomly assigned to one of four conditions in a 2 (message type: apologizing vs. blaming) $\times$ 2 (authorship label: no label vs. AI label) between-subjects design. After the consent procedures, participants read the scenario, which varied according to condition. Then they read the message from John. Following this they provided subjective probabilities in response to two questions: ``Would a \textit{warm} person ever communicate the way that John did?'' and ``Would a \textit{cold} person ever communicate the way that John did?''. Participants responded to each question on a 9-point scale, ranging from (1) \textit{extremely unlikely} to (9) \textit{extremely likely}, with (5) \textit{moderately likely} at the midpoint. Finally, they provided information about age and gender.

\subsection{Results}
\subsubsection{Message Diagnosticity:} We calculated the signal-validity score of each message for the trait category of warm using the previously described formula (see the Trait-Behavior Probability Scale section). \textbf{A score close to 0.5 indicates that the message was \textit{low diagnosticity}}: it was not thought to discriminate between warmth and coldness. \textbf{Scores above 0.5 indicate the message was more associated with warmth than coldness, while scores below 0.5 indicated the opposite. Both of these are \textit{high in diagnosticity}, just for opposite ends of the warmth spectrum}. By construction, the signal-validity statistic for warmth equals 1 minus the signal-validity statistic for coldness. For example, if an apology message has a signal-validity score of 0.8 for warmth, then its signal-validity score for coldness is automatically 0.2. Hence, for simplicity, we only report analysis on each message's signal-validity for the warmth trait.

\begin{figure}[t]
    \centering
    \includegraphics[width=\columnwidth]{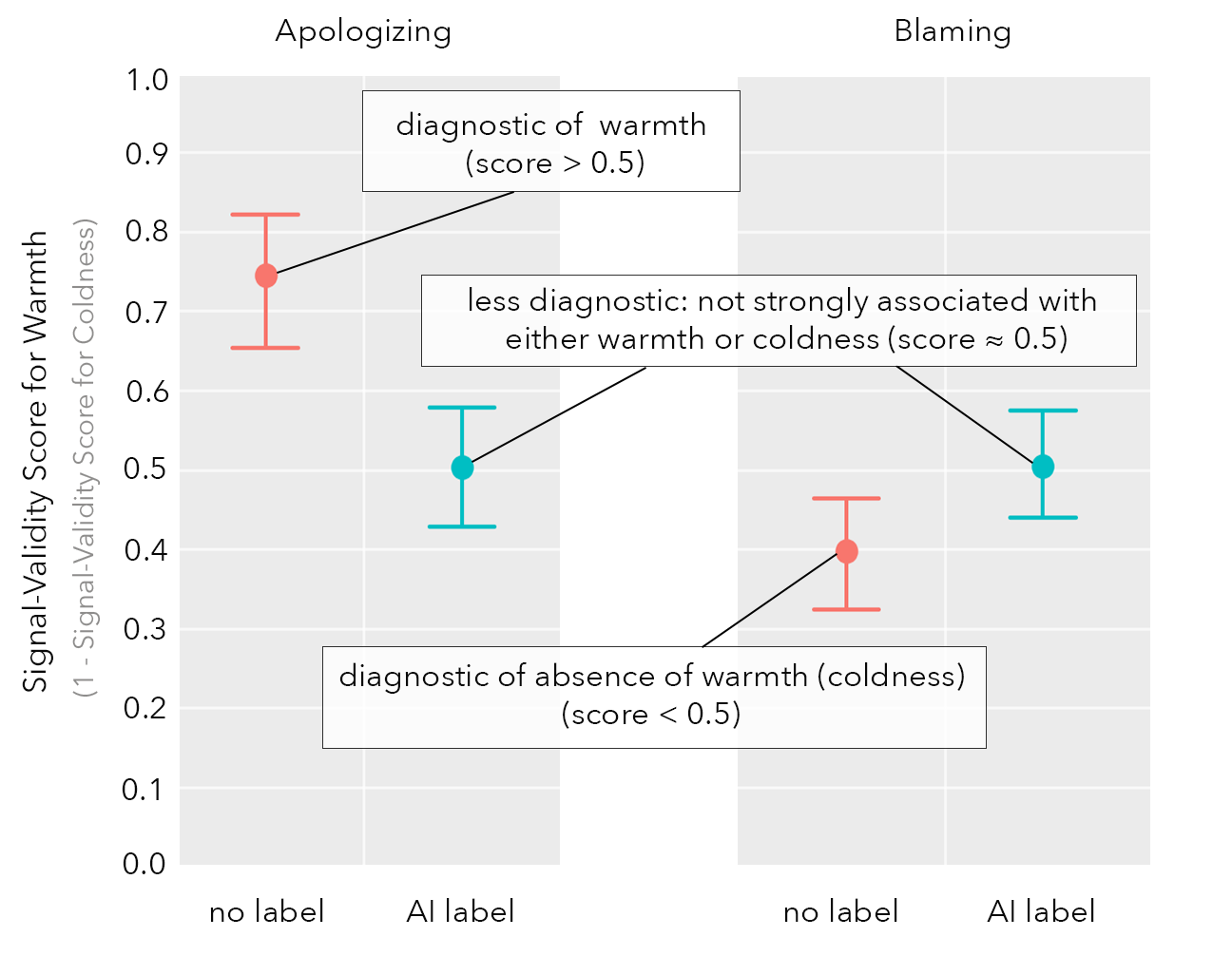}
    \caption{Study 1 shows that AI labels make both apology messages and blame messages less diagnostic (results in signal-validity scores closer to 0.5). In absence of AI labels, apology messages are diagnostic of warmth, and blame messages are diagnostic of an absence of warmth (diagnostic of coldness). Error bars are 95\% CIs.} 
    \label{diagnosticity}
\end{figure}

Figure \ref{diagnosticity} shows the means of signal-validity scores of each message corresponding to the trait category of warm. In absence of AI labels, apologizing was viewed as more diagnostic of warmth ($\text{Mean score} = 0.75$, $SD = 0.24$), and blaming was viewed as more diagnostic of absence of warmth (or more diagnostic of coldness) ($\text{Mean score} = 0.39$, $SD = 0.20$). The presence of AI labels, moves the validity score of both blaming ($\text{Mean score} = 0.50$, $SD = 0.19$) and apologizing ($\text{Mean score} = 0.50$, $SD = 0.21$) closer to 0.5, suggesting a drop in diagnosticity.

We analyzed the signal-validity score using a two-way analysis of variance (ANOVA). Signal-validity score (corresponding to the trait category warm) was the dependent variable, message type was the first factor (2 levels: apologizing, blaming), and authorship label was the second factor (2 levels: no label, AI label). The ANOVA revealed a significant interaction effect ($F(1, 115) = 20.10$, $p< 0.001$, $\eta^2 = 0.15$). We also found a large main effect of message type ($F(1, 115) = 19.83$, $p< 0.001$, $\eta^2 = 0.15$). This main effect is as expected: the signal-validity score for warmth was lower for blaming messages ($\text{Mean score} = 0.45$, $SD = 0.20$) than apologizing messages ($\text{Mean score} = 0.62$, $SD = 0.25$). Put simply, apologies are more strongly associated with warmth than blames. The main effect of authorship label was small and not significant ($F(1, 115) = 2.98$, $p = 0.09$, $\eta^2 = 0.03$). That the effect of AI labels is primarily an interaction effect (rather than a main effect) is consistent with the idea that the directional effect of AI labels depends on message type: as shown in Figure \ref{diagnosticity}, for blames it \textit{increases} the validity score for warmth and for apologies it \textit{decreases} the validity score for warmth, ultimately reducing the diagnosticity of each message (moving the score closer to 0.5). A post hoc Tukey test revealed that the difference in validity score between AI-labeled and unlabeled apologies was significant ($p < 0.001$), but the difference between AI-labeled and unlabeled blames was not ($p = 0.21$).

\subsection{Discussion}
The unlabeled apology message was associated more strongly with warmth than with coldness, whereas the unlabeled blame message was more strongly associated with coldness than with warmth. However, AI labels weakened these associations in \textit{both} cases: the AI-assisted apology and the AI-assisted blame both had signal-validity scores closer to 0.5. Taken together, these results provide preliminary evidence that appearance of AI assistance makes messages less diagnostic. 

Based on these findings, we suggest that the lower diagnosticity of AI-labeled messages may explain patterns observed in prior work~\cite{jakesch2019ai, hohenstein2020ai, hohenstein2023artificial, liu2022will, purcell2023fears}. For warmth-signalling messages like apologies, lower diagnosticity should produce drops in warmth judgments of the sender: when evidence of warmth is weaker, recipients ought to make more conservative judgments, rating the sender as less warm than they would based on stronger evidence. But, by the same token, for coldness-signalling messages like blames, lower diagnosticity should \textit{prevent} such drops---because the message becomes less strongly associated with coldness (and more strongly associated with warmth). This second prediction has not been previously tested. Testing it would also clarify whether the primary effect of AI labels is one of lowering diagnosticity or whether the effect is categorically negative: if the effect were categorically negative, we would expect AI labels to lower warmth ratings even for brags and blames. Therefore, in Study 2, we tested these predictions. A secondary objective of Study 2 was to further investigate why people instinctively view AI-labeled messages as less diagnostic. 

\section{Study 2}
The second study used the same scenario method as before, but we expanded the set of stimuli to also include thanking and bragging messages, resulting in two warmth-signalling messages (thanking, apologizing) and two coldness-signalling messages (bragging, blaming). Thus, the study featured a 4 (message type: thanking vs. apologizing vs. bragging vs. blaming) $\times$ 2 (authorship label: AI label vs. no label) between-subjects design. We included a measure for perceived warmth of the sender (for our confirmatory analysis), as well as exploratory measures to further understand the process through which people judge the sender's warmth. The main predictions being tested were that: (1) for thanking and apologizing messages, senders will be perceived as less warm when messages carry AI labels compared to no labels; and (2) for blaming and bragging messages, senders will not be perceived as less warm when messages carry AI labels compared to no labels. The study was preregistered. (Preregistration can be accessed at: \verb|https://aspredicted.org/9QX_P8F|\footnote{For brevity, we do not report on exploratory variables that were collected for different analytical purposes unrelated to the hypotheses tested.})

\subsection{Method}

\subsubsection{Participants: }
We aimed for 35 participants in each of the four conditions to detect an effect size of $f \geq 0.25$ (powered at 95\% and a significance level of 0.05). (We expected a medium effect size based on a pilot study with $80$ participants.) Again, using Prolific, we recruited 290 participants. 9 participants did not complete the survey and 1 failed the attention check, leaving 280 valid responses for analysis (33-36 in each condition; mean age = 38; 132 women and 148 men). Participants received \$2 for the 10 minute survey.

\subsubsection{Stimulus Development:} We used a scenario setup similar to Study 1. However, in Study 2, we also included thanks and brags in addition to the apologies and blames considered in Study 1. Thanks and apologies are known to be warmth-signalling and brags and blames are known to be coldness-signalling~\cite{chaudhry2019thanking}. Again, drawing on prior work~\cite{chaudhry2019thanking}, we operationalize a thanks message as the sender ``\textit{giving} the recipient credit for a positive outcome'', and brag message as the sender ``\textit{taking} credit for a positive outcome''.

Building on the scenario from Study 1, we expanded the design by additionally varying the event outcome (harshly criticized vs. effusively praised at the community meeting). This created a factorial structure where different combinations of participant role (organizer vs. helper) and outcome (positive vs. negative) naturally warranted different message types from John to the participant: thanks (participant helped John , positive outcome), apology (John helped participant, negative outcome), brag (John helped participant, positive outcome), and blame (participant helped John, negative outcome).

The content of the apologizing and blaming messages was the same as in Study 1. The thanking and bragging message were also generated in a similar manner, prompting ChatGPT (GPT-4) with the scenario setup and asking it to generate a 3-4 sentence long, text message from John. For participants assigned to see the message with AI labels, the message was additionally appended with a label that read ``parts of this message were written with the help of AI''. (The scenario, the message content, and how they were shown to participants can be found in Supplementary Material.)  

\subsubsection{Hypotheses:}
For our study context, the predicted effects of AI labels of warmth judgments translate to the following 5 hypotheses, for which we planned to conduct (and preregistered) confirmatory tests:
\begin{itemize}
    \item \textbf{H1}: There \textbf{will be} an interaction effect between message type and authorship label on perceived warmth.
    \item \textbf{H2A}: For message type of thanking, perceived warmth \textbf{will be lower} when there is AI label than when there is no label.
    \item \textbf{H2B}: For message type of apologizing, perceived warmth \textbf{will be lower} when there is AI label than when there is no label.
    \item \textbf{H3A}: For message type of bragging, perceived warmth \textbf{will not be lower} when there is AI label than when there is no label.
    \item \textbf{H3B}: For message type of blaming, perceived warmth \textbf{will not be lower} when there is AI label than when there is no label.
\end{itemize}

\subsubsection{Procedure and Measures}
Participants were randomly assigned to one of eight conditions in a 4 (message type: thanking vs. apologizing vs. bragging vs. blaming) $\times$ 2 (authorship label: no label vs. AI label) between-subjects design. After the consent procedures, participants read the scenario, which varied according to condition. Then they read the message from John. Participants then responded to two key measures. First, they rated John's \textbf{perceived warmth} on 8 items (kind, likeable, warm, trustworthy, assertive [reverse coded], competitive [reverse coded], cold [reverse coded], unfriendly [reverse coded]) using 7-point scales (1 = \textit{not at all}, 7 = \textit{extremely}), which we averaged into a composite score ($\alpha = 0.86$). Second, they rated John's \textbf{perceived competence} on 5 items (competent, intelligent, successful, hard working, skillful) using 7-point scales (1 = \textit{not at all}, 7 = \textit{extremely}), averaged into a composite score ($\alpha = 0.95$). Participants in AI label conditions additionally responded to an open-ended question that asked: ``What do you think led John to use AI to write parts of the message?''. This allowed us to capture the \textbf{causal attributions} they made for his AI use. We included perceived competence and the question about causal attributions as exploratory measures to further investigate the process through which recipients judge the sender. Finally, participants provided information about age and gender.

\subsection{Results}
\subsubsection{Perceived Warmth}
As preregistered, we created a composite score for perceived warmth by taking the average of the corresponding items ($\alpha = 0.86$), and analyzed it using a two-way analysis of variance (ANOVA). Perceived warmth was the dependent variable, message type was the first factor (4 levels: bragging, thanking, apologizing, blaming), and authorship label was the second factor (2 levels: AI label, no label).

\begin{figure}[t]
    \centering
    \includegraphics[width=\columnwidth]{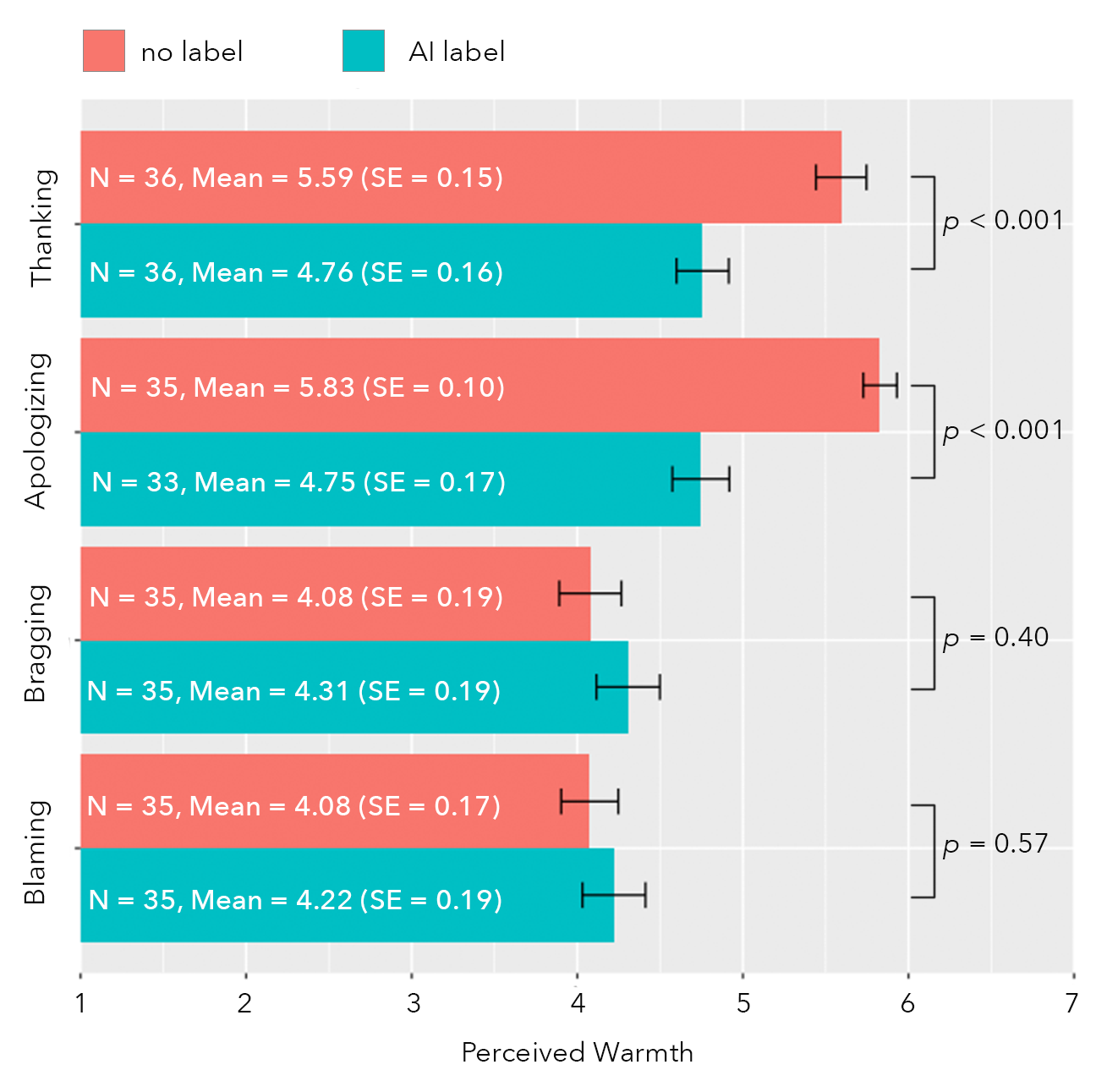}
    \caption{Study 2 shows that AI labels lead to lower warmth judgments in the case of thanking and apologizing but not for bragging and blaming. (S.E. and error bars denote standard error.)} 
    \label{warmth-result}
\end{figure}

Figure \ref{warmth-result} shows the mean perceived warmth in each study condition. The ANOVA revealed a significant interaction effect with a medium effect size ($F(3, 272) = 7.94$, $p< 0.001$, $\eta^2 = 0.08$). This provides support for \textbf{H1}. We also found a large main effect of message type ($F(3, 272) = 26.72$, $p< 0.001$, $\eta^2 = 0.23$) and a small main effect of authorship label ($F(1, 272) = 10.58$, $p< 0.01$, $\eta^2 = 0.04$). 

Further, as preregistered, we carried out four planned comparisons, corresponding to each of the four remaining hypotheses. Each planned comparison (and each remaining hypothesis) considered a single message type and contrasted the group that saw the message type with AI label to the group that saw the same message type but with no label. For each planned comparison, we used an independent samples t-test\footnote{Analyzing the contrasts via a post hoc Tukey test also leads to the same findings.}. This returns valid p-values~\cite{seltman2012experimental, midway2020comparing}, without requiring corrections, because the comparisons are: (1) planned, (2) orthogonal (each group appears in only one comparison), and (3) fewer in number than the total degrees of freedom (seven).  

For message type of thanking (Figure \ref{warmth-result}), John's perceived warmth was significantly lower ($t(69.92) = -3.81$, $p < 0.001$) among participants who saw the message with AI label ($M= 4.76$, $SD = 0.95$) compared to participants who saw the message with no label ($M = 5.59$, $SD = 0.91$). This supports \textbf{H2A}. 

Similarly, for message type of apologizing (Figure \ref{warmth-result}), John's perceived warmth was significantly lower ($t(52.20) = -5.42$, $p < 0.001$) among participants who saw the message with AI label ($M= 4.75$, $SD = 0.99$) compared to participants who saw the message with no label ($M = 5.83$, $SD = 0.60$). This supports \textbf{H2B}. 

For message type of bragging (Figure \ref{warmth-result}), there was no significant difference in John's perceived warmth ($t(67.97) = 0.85$, $p = 0.40$) across authorship label, although the average rating was higher among participants with AI label ($M= 4.31$, $SD = 1.13$) than among participants with no label ($M = 4.08$, $SD = 1.11$). This supports \textbf{H3A}.

Finally, for message type of blaming (Figure \ref{warmth-result}), there was no significant difference in John's perceived warmth ($t(67.36) = 0.57$, $p = 0.57$) across authorship label, although the average rating was higher among participants with AI label ($M = 4.22$, $SD = 1.13$) than among participants with no label ($M = 4.08$, $SD = 1.02$). This supports \textbf{H3B}.

\subsubsection{Perceived Competence}
We calculated a composite score ($\alpha = 0.95$) for perceived competence and analyzed it using a two-way ANOVA with perceived competence as the dependent variable, message type as the first factor and authorship label as the second factor. We find a main effect of authorship label ($F(1, 272) = 15.32$, $p< 0.001$, $\eta^2 = 0.05$) and a main effect of message type ($F(3, 272) = 9.74$, $p< 0.001$, $\eta^2 = 0.10$). The interaction effect was not significant and small ($F(3, 272) = 1.78$, $p = 0.15$, $\eta^2 = 0.02$). 

A post hoc Tukey test showed that John's competence was judged significantly lower ($p < 0.001$) by groups that saw AI label ($M = 4.09$, $SD = 1.37$) as compared to groups that saw no label ($M = 4.70$, $SD = 1.35$). We develop this observation further in the discussion, where we integrate it with our causal attribution analysis.

\subsubsection{Causal Attributions}
\begin{table*}[t]
\centering
\renewcommand{\arraystretch}{1.5}
\small
\begin{tabular}{@{}p{4.5cm} p{1.5cm} p{4cm} p{5.5cm}@{}}
\toprule
\textbf{Code \textit{(Group)}} & \textbf{Frequency} & \textbf{Description} & \textbf{Quote} \\
\midrule
Help Verbalizing \textit{(Competence)} & 73 & Use of AI to help find the right words to express oneself. & ``Probably just because he wanted to say something clearly or he couldn't figure out the right words to use.'' \\
\midrule
Sensitive Topic \textit{(Situation)} & 30 & Use of AI to help manage emotional reactions of the sender or recipient. & ``He was struggling with how to convey his emotions on a tough subject.'' \\
\midrule
Time \textit{(Situation)} & 22 & Use of AI to save time. & ``I think it was more of a time issue. John wanted to say 'thanks,' but didn't really have time to compose a thoughtful message.'' \\
\midrule
Laziness \textit{(Competence)} & 11 & Use of AI due to laziness. & ``I'd assume the use of AI was due to general laziness and a dislike of writing in general, rather than a personal slight directed toward myself.'' \\
\midrule
Character Judgment \textit{(Warmth)} & 9 & Use of AI due to a disregard for one's communication partner. & ``Because he didn't really want to give the credit to his helper.'' \\
\bottomrule
\end{tabular}
\caption{Participants' causal attributions for why John used AI to write a message.}
\label{tab:reasons}
\end{table*}
At the end of the survey, participants in the indication of AI use conditions were asked: ``What do you think led John to use AI to write parts of the message?'' There were a total of 139 responses (corresponding to the 139 participants across the indication of AI use conditions). These responses were inductively coded in two cycles to identify participants' interpretations of John's AI use, focused on structural and descriptive codes \cite{saldana2021coding}. This resulted in five codes (see Table \ref{tab:reasons}): \textit{help verbalizing}, \textit{sensitive topic}, \textit{time}, \textit{laziness}, and \textit{character judgment}. Some participant responses received two tags, as participants explicitly mentioned multiple possibilities (ex. \textit{``He either needed help to make it sound presentable or was simply too lazy to care''} was coded as \textit{help verbalizing} and \textit{lazy}.) In this way, the 139 responses resulted in a total of 145 tags.
\label{qual}
As a final step, after the inductive process above, we used a deductive process to group the codes based on whether they involve: (1) attributions to John's \textit{warmth} (whether John intends good or ill) (2) attributions to John's \textit{competence} (John's ability to act on those intentions) or (3) attributions to John's \textit{situation}. Theories of attribution~\cite{ross1977intuitive, gilbert1995correspondence} (inferring the causes of someone's behavior) often describe attributions in terms of disposition and situation. The deductive grouping of attributions to warmth and competence, brings together responses where participants attributed John's use of AI to his disposition. Meanwhile, deductive grouping of attributions to situation brings together responses where participants attributed John's use of AI to his situation. Groupings are shown alongside the code in Table \ref{tab:reasons}. The warmth group included one code (\textit{character judgment}), the competence group included two codes (\textit{help verbalizing} and \textit{laziness}) and the situation group included two codes (\textit{sensitive topic} and \textit{time}).

Based on this deductive grouping, we found that John's use of AI was most often attributed to competence-related factors: 84 of the 145 tags included codes from the competence group (help with verbalizing and laziness). Another 52 tags attributed AI use to situational factors, such as the sensitivity of the topic or time constraints John likely faced. Only rarely did participants interpret AI use as reflecting John's poor character or lack of warmth (9 tags).

\subsection{Discussion}
Study 2 provides empirical support for the signal diagnosticity explanation. AI labels led to lower warmth judgments in the case of thanking and apologizing but not for bragging and blaming. These results align with a dampening rather than blanket negative effect: a categorically negative view of AI assistance would predict lower warmth ratings for all messages, including brags and blames

The qualitative analysis of causal attributions provides additional support for the signal diagnosticity explanation by revealing how participants interpret AI use. When asked why John used AI assistance, the most frequent attribution was ``help verbalizing'' (73 out of 145 tags), where participants believed John needed AI to find the right words to express himself. This aligns with the signal diagnosticity interpretation: participants aren't concluding that John lacks the underlying trait entirely, but rather that he possesses it without sufficient intensity or skill to articulate it independently. As one participant noted, John ``wanted to say something clearly or he couldn't figure out the right words to use.'' This interpretation suggests that AI assistance signals a gap between having a sentiment and being able to express it effectively, making the message less diagnostic of the sender's true character. The predominance of competence-related attributions (84 tags) over warmth-related ones (9 tags) further supports this view---participants see AI use as reflecting limitations in expressive ability rather than absence of the underlying feeling. This interpretation is reinforced by the finding that AI labels reduced perceived competence ratings, suggesting participants view AI assistance predominantly as evidence of reduced communicative skill.

\section{General Discussion}
Across two studies, we found support for the \textit{signal diagnosticity explanation}: that AI labels result in messages being viewed as less diagnostic of the sender's moral character. We now turn to discussing how our work sheds light on previously reported phenomena, including how AIMC serves as a moral crumple zone~\cite{hohenstein2020ai} and affects interpersonal trust~\cite{jakesch2019ai}. Finally, we discuss design opportunities, outline limitations of our work, and identify directions for future research.

\subsection{AI as a Moral Dampener Rather Than a Moral Disqualifier}
Rather than AI labels serving as categorical disqualifiers of moral character (marking senders as definitively cold or calculating), they seem to function as dampeners: AI labels don't just reduce the positive impact of warm messages, they also limit the negative impact of cold ones. The mere presence of AI assistance acts as a social buffer, absorbing interpersonal damage when conversations go poorly

This sheds light on the psychological basis for the previous observation of AIMC as a moral ``crumple zone''~\cite{hohenstein2020ai}—a concept borrowed from automotive safety where certain parts of a car are designed to absorb impact during crashes, protecting the occupants. This previous work found that when recipients knew that AI was involved in crafting messages, they held the sender less responsible for negative conversational outcomes, prompting the researchers to conclude that AI mediation can strengthen relationships between communicators.

Our work provides support for this perspective but also reveals a more complex picture. The very feature that makes AI-assisted criticisms feel survivable makes AI-assisted thank you notes feel hollow. While dampening can prevent relationship damage in negative interactions, it can simultaneously undermine relationship building in positive ones.

\subsection{In Search for Honest Signals}
Previous research identified the \textit{``Replicant Effect''}~\cite{jakesch2019ai}, where individuals merely suspected of relying on AI assistance for warmth-signalling messages like profile greetings were judged less warm. Our work shows that even when people have complete certainty about AI involvement (through explicit labels), the depression in warmth assessment persists.

People seem to inherently view unintermediated communication as a more ``honest signal''~\cite{donath2007signal1, donath2007signal2, zahavi1975mate}---a signal that more reliably indicates underlying traits because it is costlier or more difficult to fake~\cite{zahavi1975mate}.  When someone expresses gratitude or remorse without assistance, many of us may instinctively read it as evidence that these feelings are strong enough to flow naturally into words. Here, the cognitive and emotional effort required to translate feeling into language serves to authenticate the emotion itself. By reducing or eliminating this effort, AI assistance strips away these authentication markers and recipients are left unable to distinguish between someone who deeply feels an emotion and someone who merely recognizes they should express it.

As AI assistance becomes ubiquitous, we risk creating a communicative race---where everyone must escalate their expressions of warmth just to achieve the same interpersonal impact that unassisted communication once delivered. This could result in paradoxical outcomes: as our messages become more polished and eloquent through AI, they simultaneously become less meaningful and less trusted. The handwritten note gains value precisely because it cannot be automated; the stuttered apology may repair relationships more effectively than the eloquent one. This suggests that preserving spaces for effortful expression isn't just nostalgic sentiment but a functional necessity for maintaining trust in human relationships.

\subsection{Designs to Reduce Attributional Ambiguity}
Rather than abandoning effortful expression entirely, we suggest that communication technologies could be designed to preserve and highlight markers of human investment~\cite{he2025contributions}. There is an opportunity to investigate whether AI tools that reveal the sender's emotional and cognitive investment could help enhance relationship building and emotional connection. Future work could test whether effort-revealing designs~\cite{kelly2017demanding, zhang2022auggie} that highlight personal investment---such as showing that someone struggled to find the right words or spent significant time crafting their response---generate stronger feelings of warmth and appreciation than blanket AI-assistance labels. Understanding how these design choices affect relational satisfaction and emotional bonding could inform the development of AI communication tools that enhance rather than diminish human connection.

\subsection{Limitations}
Our study focusing on hypothetical scenarios offered experimental evidence of the signal diagnosticity explanation. However, this experiment did not allow us to test whether there are behavioral consequences of these effects. For instance, if someone is known to have used AI when blaming, is the recipient's response different than if there was no disclosure of AI use? Future work that investigates these message types in the context of live interactions can help identify the behavioral consequences. 

Our focus in this work was on the recipient's perspective, rather than the sender's perspective. We were unable to investigate, for instance, whether senders would choose to use AI assistance for some message types over others. We were also unable to investigate whether the use of AI affected whether senders \textit{felt} like they bragged, blamed, apologized, or thanked. Recent studies suggest that use of AI assistance can also affect the sender's self-perceptions about their agency and control~\cite{kadoma2024role, hwang202580}. Future investigation that considers the sender's perspective can advance our initial findings to strengthen the conceptual foundations of AI-mediated communication. 

Finally, our work did not consider the ways in which actually composing a message with an AI tool can itself influence the sender's message~\cite{arnold2020predictive, jakesch2023co}. Consistent with prior work~\cite{liu2022will, jakesch2019ai, weiss2022effects}, we opt for a controlled test of our hypotheses by keeping the content of the message constant across the indication of AI use and no indication of AI use conditions. However, prior work has documented how biases in generative AI (towards certain generation lengths or towards certain opinions) can influence what people write~\cite{arnold2020predictive} and think~\cite{jakesch2023co}. Future work is needed to understand how the technical underpinnings of current systems might interact with the effects we study.

\section{Conclusion}
When people see ``written with AI,'' they do not conclude the sender is cold or calculating; they conclude the sender possesses their expressed sentiment, but the degree of the sentiment is obfuscated by AI. This dampening effect means AI labels weaken both positive and negative character impressions: AI-assisted apologies seem less warm, but AI-assisted blame seems less cold. These reactions provide credence to the idea that AI labels reduce the diagnosticity of messages, and help explain previously observed phenomena in AIMC. Our work offers a foundation for studying how perceptions of communication partners, rather than the content they share, change with AI integration.

\section{Reflection on Ethics and Societal Impact}

\subsection{Positionality and Adverse Impact Statement}
Our US-based team comprises individuals with diverse academic backgrounds, including human computer interaction (HCI), responsible AI, and psychology. Our backgrounds have informed how we frame and pursue our investigations.

Our research participants were all US-based, and so, they may share similar cultural assumptions about the relationship between expressive ability and moral judgments. The signal diagnosticity framework we employ reflects particular cultural values about individual agency and the importance of unmediated self-expression that may not generalize across all cultural contexts.

We also recognize that our focus on warmth and competence as primary dimensions of moral character reflects established Western psychological frameworks, which may not capture the full complexity of how different communities evaluate moral character or the role of technological mediation in social relationships.

Our goal has been to advance academic and public debates about AI's societal impact, while maintaining scientific rigor in our methodology and interpretation of results. Nevertheless, we invite readers to approach our findings with awareness of these contextual limitations.

\subsection{Ethical Considerations}
Both studies were approved by our university's Institutional Review Board. Our study included a minor psychological risk, given that some participants were exposed to a psychologically threatening (i.e. blaming) message. This risk was offset by the fictitious nature of the scenario, and further offset by the fact that participants were asked to reflect on the sender's motivation. Reflecting on an aggressor's motivation can help dampen negative psychological impacts \cite{gyurak2011explicit}.

\section{Acknowledgments}
We thank our anonymous reviewers for their feedback on the paper. We also thank Chinmay Kulkarni for insightful discussions and valuable feedback that helped shape this work.

\bibliography{main}

\newpage
\appendix
\begin{strip}
    \centering
    {\Large\bfseries Supplementary Material}
    \par\vspace{0.5em}
\end{strip}
\section{Materials for Study 1}
\label{study-details}
\subsection{Scenario Template}
\label{scenario}
Different configurations of the scenario manipulate the participant's role (originator v/s receiver of help, differences in scenarios indicated in bold) to create warrant for the other character (`John') to either apologize to or blame the participant. The scenario template is as follows:

\begin{tcolorbox}[colback=gray!10, colframe=black, width=\columnwidth, boxrule=0.1mm, arc=2mm, auto outer arc]
\begin{center}
\textit{You and John}
\end{center}
\textit{Imagine that you and your friend John are members of a local community organization that plans events to bring the neighborhood together. You both have been involved in this group for a while, and your bond has grown strong over time.}
\begin{center}
\textit{The Task}
\end{center}
\textit{The community organization is planning a large neighborhood fair, and \textbf{you were/John was} given the responsibility of organizing the event. \textbf{You asked John/John asks you} for \textbf{his/your} input, and \textbf{he/you} spent a lot of time helping \textbf{you/him} plan the event, suggesting some significant changes, which \textbf{you/he} accepted.}
\begin{center}
\textit{The Outcome}
\end{center}
\textit{During the next community meeting, a majority of attendees harshly criticized \textbf{your/John’s} plan, particularly highlighting the parts that \textbf{John is}/ \textbf{you are} primarily responsible for.}
\end{tcolorbox}
These scenarios create a context where either an apology or blame message from John is situationally warranted, allowing us to test how the presence of AI assistance labels affects participants' perceptions of John's character. An apology message from John is warranted in scenarios where the participant is the primary organizer of the event and John assisted them, while a blame message from John is warranted in scenarios where John is the primary organizer and the participant assisted him.

\subsection{Message Content}
\label{messages}
For participants in AI label conditions, the message is additionally appended with an indicator that reads ``parts of this message were written with the help of AI''.
\\

{\noindent
\textbf{Apologizing:}}
\begin{tcolorbox}[colback=gray!10, colframe=black, width=\columnwidth, boxrule=0.1mm, arc=2mm, auto outer arc]
\textit{``Hey! Just wanted to say I'm really sorry about how the event planning got critiqued yesterday. I know a lot of the parts the group didn't like were the changes I suggested. I honestly thought they'd help, but I guess I messed up. Sorry again, and I hope we’re cool.''}
\end{tcolorbox}

{\noindent
\textbf{Blaming:}}
\begin{tcolorbox}[colback=gray!10, colframe=black, width=\columnwidth, boxrule=0.1mm, arc=2mm, auto outer arc]
\textit{``I'm really upset about the meeting yesterday. The harsh criticism was mostly on the parts you helped me with. I trusted your suggestions, and now I'm worried.''}
\end{tcolorbox}

\subsection{Message visual}
\begin{figure}[h]
    \centering
    \includegraphics[width=0.6\columnwidth]{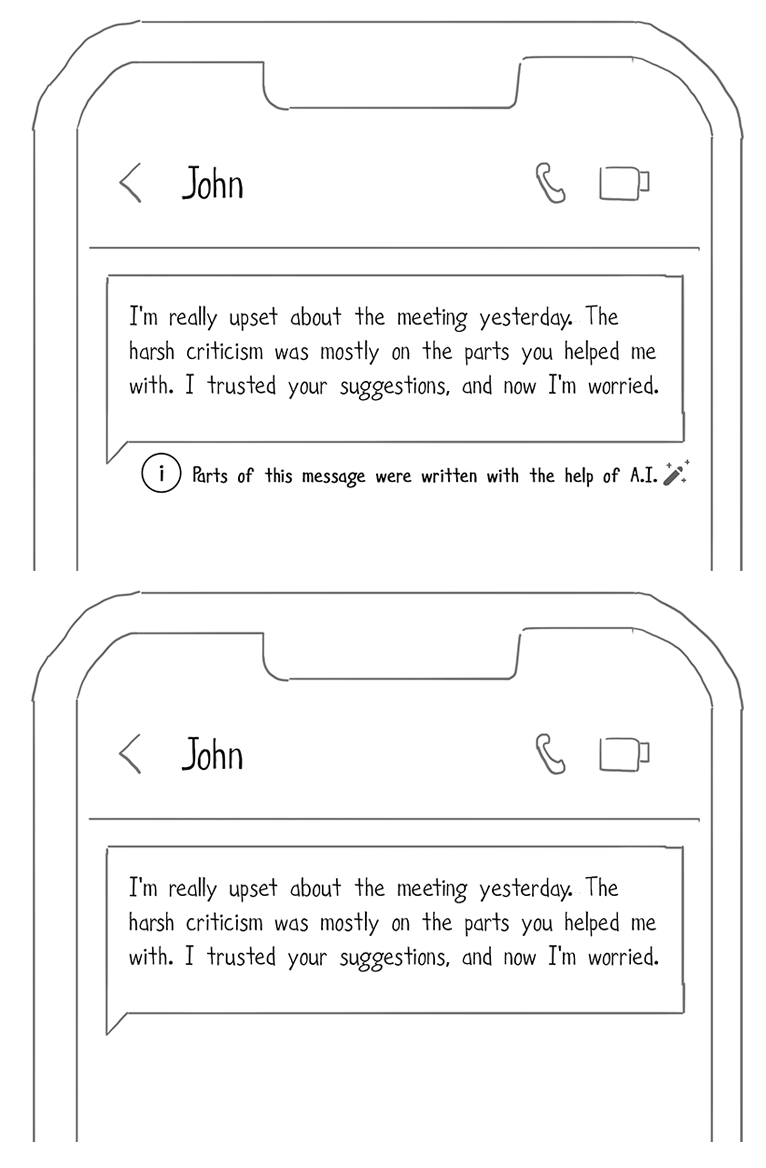}
\end{figure}
Examples of how the messages are presented to participants in the study. Top represents the blame with AI label and bottom is the blame without AI label. Each participant only sees one message.


\section{Materials for Study 2}

\subsection{Scenario Template}
\label{scenario}
In Study 2, we vary both the participant's role (either the originator who provides John help or the receiver who receives John's help, indicated in bold) and the outcome (positive or negative, indicated with underline in the text). This creates natural contexts for four different message types from John: thanking (positive outcome, participant as originator), blaming (negative outcome, participant as originator), bragging (positive outcome, participant as receiver), and apologizing (negative outcome, participant as receiver). Each scenario establishes a situation where one of these message types would be socially warranted.

\begin{tcolorbox}[colback=gray!10, colframe=black, width=\columnwidth, boxrule=0.1mm, arc=2mm, auto outer arc]
\begin{center}
\textit{You and John}
\end{center}
\textit{Imagine that you and your friend John are members of a local community organization that plans events to bring the neighborhood together. You both have been involved in this group for a while, and your bond has grown strong over time.}
\begin{center}
\textit{The Task}
\end{center}
\textit{The community organization is planning a large neighborhood fair, and \textbf{you were/John was} given the responsibility of organizing the event. \textbf{You asked John/John asks you} for \textbf{his/your} input, and \textbf{he/you} spent a lot of time helping \textbf{you/him} plan the event, suggesting some significant changes, which \textbf{you/he} accepted.}
\begin{center}
\textit{The Outcome}
\end{center}
\textit{During the next community meeting, a majority of attendees [\underline{effusively praised}/ \underline{harshly criticized}] \textbf{your/John’s} plan, particularly highlighting the parts that \textbf{John is}/ \textbf{you are} primarily responsible for.}
\end{tcolorbox}

\subsection{Message Content}
\label{messages}
For participants in AI label conditions, the message is additionally appended with an indicator that reads ``parts this message were written with the help of AI''.
\\

{\noindent
\textbf{Thanking:}}
\begin{tcolorbox}[colback=gray!10, colframe=black, width=\columnwidth, boxrule=0.1mm, arc=2mm, auto outer arc]
\textit{``Hey! Just wanted to say a huge thanks for all your help with the event planning. Your feedback was spot on, and I really think it turned out great because of you. I really appreciate you taking the time to go through it so thoroughly.''}
\end{tcolorbox}

{\noindent
\textbf{Apologizing:}}
\begin{tcolorbox}[colback=gray!10, colframe=black, width=\columnwidth, boxrule=0.1mm, arc=2mm, auto outer arc]
\textit{``Hey! Just wanted to say I'm really sorry about how the event planning got critiqued yesterday. I know a lot of the parts the group didn't like were the changes I suggested. I honestly thought they'd help, but I guess I messed up. Sorry again, and I hope we’re cool.''}
\end{tcolorbox}

{\noindent
\textbf{Blaming:}}
\begin{tcolorbox}[colback=gray!10, colframe=black, width=\columnwidth, boxrule=0.1mm, arc=2mm, auto outer arc]
\textit{``I'm really upset about the meeting yesterday. The harsh criticism was mostly on the parts you helped me with. I trusted your suggestions, and now I'm worried.''}
\end{tcolorbox}

{\noindent
\textbf{Bragging:}}
\begin{tcolorbox}[colback=gray!10, colframe=black, width=\columnwidth, boxrule=0.1mm, arc=2mm, auto outer arc]
\textit{``Hey! Just wanted to say I'm glad the event planning was well-received. I put a lot of effort into those changes, and it felt great to see them get praised. Feels good to be the mastermind behind the best parts of the plan!''}
\end{tcolorbox}

\subsection{Message visual}

Messages are displayed the same way as in Study 1.

\end{document}